\documentclass[11pt]{article}
\usepackage{amscd,amsmath,amssymb,latexsym,bbm,graphicx}

\catcode`@=11
\makeatletter\renewcommand{\section}{\@startsection
{section}{1}{\z@}{-3.5ex plus -1ex minus
    -.2ex}{2.3ex plus .2ex}{\large\bf }}

\makeatletter\renewcommand{\subsection}{\@startsection{subsection}{2}{\z@}{-3.25ex
plus -1ex minus
   -.2ex}{1.5ex plus .2ex}{\bf }}

\numberwithin{equation}{section}
\newcounter{saveeqn}

\catcode`@=12

\def\b{\beta}

\def\h{\eta}
\def\th{\theta}

\def\la{\lambda}
\def\m{\mu}
\def\n{\nu}
\def\r{\rho}
\def\s{\sigma}

\def\1{\bar 1}
\def\2{\bar 2}
\def\3{\bar 3}

\newcommand{\yb}{\bar{y}}

\newcommand{\zb}{\bar{z}}

\newcommand{\C}{\mathbb C}
\newcommand{\R}{\mathbb R}
\newcommand{\Z}{\mathbb Z}

\newcommand{\Zcal}{{\cal Z}}
\newcommand{\Acal}{{\cal A}}

\newcommand{\Fcal}{{\cal F}}
\newcommand{\Ecal}{{\cal E}}
\newcommand{\J}{{\cal J}}
\newcommand{\Ucal}{{\cal U}}

\def\im{\mbox{i}}
\def\pa{\mbox{$\partial$}}
\def\diff{\mbox{d}}
\def\tr{{\rm tr}}
\def\sfrac#1#2{{\textstyle\frac{#1}{#2}}}
\def\>{\rangle}
\def\<{\langle}
\def\+{\dagger}
\def\={\ =\ }
\def\and{\quad\textrm{and}\quad}
\def\with{\quad\textrm{with}\quad}

\begin{document}

\begin{titlepage}
\setcounter{page}{0}
\begin{flushright}
.
\end{flushright}

\vskip 3.0cm

\begin{center}

{\Large\bf
Integrable vortex-type equations on the two-sphere
   }

\vspace{12mm}

{\large
Alexander~D.~Popov}
\\[8mm]
\noindent {\em
Bogoliubov Laboratory of Theoretical Physics, JINR\\
141980 Dubna, Moscow Region, Russia}\\
{Email: popov@theor.jinr.ru}

\vspace{12mm}

\begin{abstract}
\noindent We consider the Yang-Mills instanton equations on the four-dimensional manifold
$S^2\times \Sigma$, where $\Sigma$ is a compact Riemann surface of genus $g>1$ or its
covering space $H^2=\ $SU(1,1)/U(1). Introducing a natural ansatz for the gauge potential,
we reduce the instanton equations on $S^2\times \Sigma$ to vortex-type equations
on the sphere $S^2$. It is shown that when the scalar curvature of the manifold
$S^2\times \Sigma$ vanishes, the vortex-type equations are integrable, i.e. can be
obtained as compatibility conditions of two linear equations (Lax pair) which are written
down explicitly. Thus, the standard methods of integrable systems can be applied for
constructing their solutions. However, even if the scalar curvature of $S^2\times \Sigma$
does not vanish, the vortex equations are well defined and have solutions for any values of
the topological charge $N$. We show that any solution to the vortex equations on
$S^2$ with a fixed topological charge $N$ corresponds to a Yang-Mills instanton on
$S^2\times \Sigma$ of charge $(g-1)N$.
\end{abstract}
\end{center}
\end{titlepage}

\section{Introduction and summary}
\noindent
The Abelian Higgs model on $\R\times\R^2$ at critical value of the coupling constant
(the Bogomolny regime) admits static vortex solutions on $\R^2$~\cite{AbNO} which describe
magnetic flux tubes (vortex strings) penetrating a two-dimensional superconductor. Vortices are
important objects in modern field theory~\cite{MS} since it is believed that (electric) vortex
strings play an important role in the confinement of quarks. Their stability is ensured by
topology~\cite{JT}. Many results known for the Abelian Higgs model were generalized to
Riemann surfaces, noncommutative spaces and to the non-Abelian case
(see e.g.~\cite{Bradlow}-\cite{MRS} and references therein).

It was shown recently that the vortex equations on a Riemann surface $\Sigma$ of genus $g$ have a
Lax pair representation if $g>1$ and do not have it for $g=0,1$~\cite{Popov}. This was done by
using the correspondence between vortices on $\Sigma$ and SU(2)-equivariant\footnote{This means a
generalized SU(2)-invariance, i.e. invariance under space-time transformations up to gauge
transformations~\cite{FM}.} instantons on the four-manifold $\Sigma\times S^2$ -- the invariance
conditions reduce the instanton equations on $\Sigma\times S^2$ to vortex equations on $\Sigma$.
Existence of a Lax pair for the reduced equations on $\Sigma$ is related with vanishing of scalar curvature
of $\Sigma\times S^2$ when this manifold becomes~\cite{BLB} a gravitational instanton.
The nonexistence of a Lax pair for vortex equations on $S^2$, $T^2$ and $\R^2$ followed from the fact that
the scalar curvature of $\Sigma\times S^2$ is non-vanishing for $\Sigma=S^2, T^2$ and $\R^2$.

In this paper, we introduce an ansatz reducing the instanton equations on $M=S^2\times\Sigma$ to vortex-type
equations not on $\Sigma$ but on $S^2$, and show that these equations are the compatibility conditions
of two linear equations (Lax pair) if the scalar curvature of $M$ vanishes,
similar to the previous $g>1$ cases~\cite{Popov}. Furthermore, the existence of solutions to the reduced
equations on $S^2$ for any topological charge $N\ge 0$ demands noncompact initial gauge group for Yang-Mills
theory on $M$ and compact gauge group of reduced Yang-Mills-Higgs theory on $S^2$. This is similar to the
case of the Hitchin equations on $S^2$ and $T^2$ obtainable as reduction of the instanton
equations~\cite{Hitchin} - smooth solutions on $S^2$ (and $T^2$) exist only if one chooses noncompact
gauge group\footnote{Yang-Mills fields with noncompact gauge groups were considered in many papers
(see e.g.~\cite{Dolan, DJ} and references therein).} in four dimensions~\cite{MJ}.

The organization of this paper is as follows. In section 2 we collect various facts concerning the geometry
of the manifold $S^2\times H^2$, where $H^2=\ $SU(1,1)/U(1) is the unit disk in the complex plane $\C$. Explicit
form of metric, Christoffel symbols etc. are written down. Then, in section 3, we introduce an
SU(1,1)-equivariant ansatz which reduces the instanton equations on $S^2\times H^2$ to Abelian
vortex-type equations on $S^2$. Solutions to these equations give solutions of the self-dual Yang-Mills
equations on $S^2\times H^2$ with the noncompact gauge group SU(1,1). Section 4 deals with integrability
properties of the introduced Abelian vortex equations. Finally, in section 5 and 6, we generalize results of
section 2-4 to the case of non-Abelian vortex-type equations on $S^2$ and instantons on manifolds
$S^2\times\Sigma$ with compact Riemann surfaces $\Sigma$. Bogomolny transformations for the Yang-Mills-Higgs
action functional is discussed and a relation between the instanton and vortex topological charges is derived.

\section{Manifold $S^2\times H^2$}

\noindent
{\bf Riemann sphere.} Consider the standard two-sphere $S^2\cong\C P^1=\,$SU(2)/U(1) of constant radius
$R_1$. In local coordinates $y=x^1+\im\,x^2, \yb=x^1-\im\,x^2$ on $\C P^1$ the metric and the volume form
read
\begin{equation}\label{2.1}
\diff s_{S^2}^2 = 2g_{y\yb}\,\diff y\,\diff \yb = \frac{4R_1^4}{(R_1^2+y\yb)^2}\,\diff y\,\diff \yb
\end{equation}
and
\begin{equation}\label{2.2}
\omega_ {S^2}= \frac{2\im\,R_1^4}{(R_1^2+y\yb)^2}\,\diff y\wedge\diff \yb = \im\, g_{y\yb}\,
\diff y\wedge\diff \yb\ ,
\end{equation}
respectively. For the nonvanishing components of the Christoffel symbols and the Ricci tensor we have
\begin{equation}\label{2.3}
\Gamma^y_{yy} = 2\,\pa_y\log\rho_1\and
\Gamma^{\yb}_{\yb\yb} = 2\,\pa_{\yb}\log\rho_1\quad\mbox{with}\quad
\rho_1^2:=g_{y\yb}\ ,
\end{equation}
\begin{equation}\label{2.4}
R_{y\yb} = -2\,\pa_y\pa_{\yb}\log\rho_1 =\frac{1}{R^2_1}\, g_{y\yb}\quad\Longrightarrow\quad
R^{}_{S^2} = 2 g^{y\yb}R_{y\yb}=\frac{2}{R^2_1}\ ,
\end{equation}
where $R^{}_{S^2}$ is the scalar curvature of $S^2$.

For the components $g_{y\yb}$ and $g^{y\yb}=1/g_{y\yb}$ we have
\begin{equation}\label{2.5}
g_{y\yb}=e^1_{y}\,e^{\1}_{\yb}\and
g^{y\yb}=e^{y}_1\,e_{\1}^{\yb}\ ,
\end{equation}
where $e^{y}_1$ and $e^{\yb}_{\1}$ are unitary (local) frame.
We introduce a basis of type (1,0) and (0,1) vector fields
\begin{equation}\label{2.6}
e_1:=e^{y}_1\,\pa_{y}\and
e_{\1}:=e^{\yb}_{\1}\,\pa_{\yb}
\end{equation}
on $S^2\cong \C P^1$. The dual basis of type (1,0) and (0,1) forms is $e_{y}^1\diff{y}$ and
$e_{\yb}^{\1}\diff{\yb}$.

\noindent
{\bf Coset space $H^2$.} Consider the symmetric space (unit disk)
\begin{equation}\label{2.7}
H^2=\mbox{SU}(1,1)/\mbox{U}(1)\ ,
\end{equation}
where SU(1,1) is a noncompact real form of the group SL(2,$\C$) with elements $h$ defined by
\begin{equation}\label{2.8}
 h^{\+}\h h=\h\quad\mbox{for}\quad\h = \begin{pmatrix}1&0\\0&-1\end{pmatrix}\ .
\end{equation}
The metric and the K\"ahler form in the coordinates $z=x^3-\im\,x^4, \zb =x^3+\im\,x^4$ on $H^2$ are given  by
\begin{equation}\label{2.9}
\diff s_{H^2}^2 = 2g_{z\zb}\,\diff z\,\diff \zb = \frac{4R_2^4}{(R_2^2-z\zb)^2}\,\diff z\,\diff \zb\ ,
\end{equation}
and
\begin{equation}\label{2.10}
\omega_ {H^2}= -\frac{2\im\,R_2^4}{(R_2^2-z\zb)^2}\,\diff z\wedge\diff \zb = -\im\, \b\wedge\bar\b\ ,
\end{equation}
where
\begin{equation}\label{2.11}
\b:=\frac{\sqrt{2}\,R_2^2\,\diff z}{R_2^2-z\zb}\and
\bar\b:=\frac{\sqrt{2}\,R_2^2\,\diff {\zb}}{R_2^2-z{\zb}}
\end{equation}
are forms on $H^2$ of type (1,0) and (0,1). These forms satisfy the equations
\begin{equation}\label{2.12}
 \diff\b = -2a\wedge\b\ ,\quad \diff\bar\b = 2a\wedge\bar\b\and
\diff a= -\frac{1}{2R^2_2}\,\b\wedge\bar\b =-\frac{\im}{2R^2_2}\,\omega_{H^2}\ .
\end{equation}
The anti-hermitian connection one-form
\begin{equation}\label{2.13}
a= \frac{1}{2(R_2^2-z\zb)}(\zb\,\diff z - z\,\diff \zb )
\end{equation}
with the curvature form $\diff a$ given in (\ref{2.12}) is an $H^2$-analog of the monopole connection
on $\C P^1$. Note that $2a$ is the Levi-Civita connection on the tangent bundle $TH^2$. The one-form $a$
is a connection on the square root $L$ of the holomorphic bundle $T^{1,0}H^2$.

The Christoffel symbols, Ricci tensor and scalar curvature for $H^2$ are
\begin{equation}\label{2.14}
\Gamma^z_{zz} = 2\,\pa_z\log\rho_2\qquad\mbox{and}\qquad
\Gamma^{\zb}_{\zb\zb} = 2\,\pa_{\zb}\log\rho_2\qquad\mbox{with}\qquad
\rho_2^2:=g_{z\zb}\ ,
\end{equation}
\begin{equation}\label{2.15}
R_{z\zb} = -2\,\pa_z\pa_{\zb}\log\rho_2 =-\frac{1}{R^2_2}\, g_{z\zb}\qquad\Longrightarrow\qquad
R^{}_{H^2} = 2 g^{z\zb}R_{z\zb}=-\frac{2}{R^2_2}\ .
\end{equation}
For (1,0) and (0,1) vector fields on $H^2$ dual to forms (\ref{2.11}) we have
\begin{equation}\label{2.16}
e_2:=e_2^z\pa_z=\rho_2^{-1}\pa_z\and e_{\2}:=e_{\2}^{\zb}\pa_{\zb}=\rho_2^{-1}\pa_{\zb}
\end{equation}
with $\rho_2$ given in (\ref{2.14}) and (\ref{2.9}).

We also consider a four-manifold $M$ given by a product of $S^2$ and $H^2$ with the product metric
\begin{equation}\label{2.17}
 \diff s^2_M =\diff s^2_{S^2} + \diff s^2_{H^2}\ .
\end{equation}
For the scalar curvature of $M=S^2\times H^2$ we have
\begin{equation}\label{2.18}
 R_M = R_{S^2} + R_{H^2} = 2\left(\frac{1}{R^2_1}-\frac{1}{R^2_2}\right)\ .
\end{equation}

\section{Vortices on $S^2$ as Yang-Mills configurations on $S^2\times H^2$}

\noindent
{\bf SU(1,1)-equivariant gauge potential.} Consider the manifold $M=S^2\times H^2$. Let $\Ecal\to M$
be an SU(1,1)-equivariant complex vector bundle of rank 2 over
$M$ with the group SU(1,1) acting trivially on $S^2$ and in the standard way by SU(1,1)-isometry on
$H^2=\,$SU(1,1)/U(1).
Let $\Acal$ be an $su(1,1)$-valued local form of SU(1,1)-equivariant connection on $\Ecal$ (cf.
\cite{PS, Popov}); it can be chosen in the form
\begin{equation}\label{3.1}
\Acal = \begin{pmatrix}\sfrac12A\otimes 1+ 1\otimes a& \sfrac{1}{\sqrt{2}}\phi\otimes\b
\\\sfrac{1}{\sqrt{2}}\bar\phi\otimes\bar\b&-\sfrac12A\otimes 1 - 1\otimes a\end{pmatrix}=\left(\sfrac12\,A
+a\right)\sigma_3 + \sfrac{1}{\sqrt{2}}\phi\b\s_+ +\sfrac{1}{\sqrt{2}}\bar\phi\bar\b\s_-\ ,
\end{equation}
where
\begin{equation}\label{3.6}
\s_3=\begin{pmatrix}1&0\\0&-1\end{pmatrix}\ ,\quad \s_+=\begin{pmatrix}0&1\\0&0\end{pmatrix}
\quad\mbox{and}\quad\s_-=\begin{pmatrix}0&0\\1&0\end{pmatrix}\ .
\end{equation}
Here, $A=A_y\diff y + A_{\bar y}\diff\bar y$ is an Abelian connection on a (Hermitian) complex
line bundle $E$ over $\C P^1\cong S^2$, $a$ is the  connection (\ref{2.13}) on the complex line
bundle $L$ over $H^2$, $\phi$ is a section of the bundle $E$, $\bar\phi$ is its
complex conjugate and forms $\b, \bar\b$ on $H^2$ are given in (\ref{2.11}).
In local complex coordinates $y,\yb$ on $\C P^1$ we have $A=A(y,\yb )$ and $\phi =\phi (y, \yb)$.

\bigskip

\noindent
{\bf Field strength tensor.} In local coordinates on $S^2\times H^2$ the calculation of the curvature
$\Fcal$ for $\Acal$ of the form (\ref{3.1}) yields
\begin{equation}\label{3.2}
\Fcal = \diff\Acal +\Acal\wedge\Acal =\begin{pmatrix}\sfrac12F -
\sfrac12\left(\sfrac{1}{R_2^2}-\phi\bar\phi \right)\b\wedge\bar\b&
\sfrac{1}{\sqrt{2}}(\diff\phi + A\phi )\wedge\b
\\[6pt]
\sfrac{1}{\sqrt{2}}(\diff\bar\phi - A\bar\phi )\wedge\bar\b&-\sfrac12F +
\sfrac12\left(\sfrac{1}{R^2_2}-\phi\bar\phi \right)\b\wedge\bar\b\end{pmatrix}\\[6pt]
\end{equation}
\begin{equation}\nonumber
=\Fcal_{y\yb}\,\diff y\wedge\diff\yb + \Fcal_{yz}\,\diff y\wedge\diff z +
\Fcal_{y\zb}\,\diff y\wedge\diff\zb + \Fcal_{\yb z}\,\diff \yb\wedge\diff z
+\Fcal_{\yb\zb}\,\diff \yb\wedge\diff\zb +\Fcal_{z\zb}\,\diff z\wedge\diff\zb
\end{equation}
with the non-vanishing field strength components
\begin{equation}\label{3.3}
\Fcal_{y\yb}=\frac{1}{2} F_{y\yb}\,\s_3\ ,\quad
\Fcal_{z\zb}=-\frac{1}{2} g_{z\zb}\left(\frac{1}{R_2^2}-\phi\bar\phi \right)\,\s_3\ ,
\end{equation}
\begin{equation}\label{3.4}
\Fcal_{\yb z}=\frac{\rho_2}{\sqrt{2}}\,(\pa_{\yb}\phi + A_{\yb}\phi )\,\s_+
\ ,\quad
\Fcal_{yz}=\frac{\rho_2}{\sqrt{2}}\,(\pa_{y}\phi + A_{y}\phi )\,\s_+\ ,
\end{equation}
\begin{equation}\label{3.5}
\Fcal_{y\zb}=\frac{\rho_2}{\sqrt{2}}\,(\pa_{y}\bar\phi - A_{y}\bar\phi )\,\s_-
\ ,\quad
\Fcal_{\yb \zb}=\frac{\rho_2}{\sqrt{2}}\,(\pa_{\yb}\bar{\phi} - A_{\yb}\bar{\phi} )\,\s_-
\ .
\end{equation}
In (\ref{3.3}) we have defined $F=\diff A=F_{y\yb}\,\diff y\wedge \diff\yb=(\pa_y A_{\yb}
-\pa_{\yb}A_y)\,\diff y\wedge \diff\yb$ for $A=A_y\diff y+A_{\yb}\diff\yb$.

\bigskip
\noindent
{\bf Vortex equations on $S^2$}. Let us consider the self-dual Yang-Mills equations $\ast\Fcal =\Fcal$ on
$S^2\times H^2$, where $\ast$ is the Hodge operator. In local coordinates these equations have the form
\begin{equation}\label{3.7}
\Fcal_{\yb\zb}=0=(\Fcal_{yz})^\+\qquad\mbox{and}\qquad
g^{y\yb}\Fcal_{y\yb} + g^{z\zb}\Fcal_{z\zb}=0\ .
\end{equation}
Substitution of (\ref{3.3})-(\ref{3.5}) into (\ref{3.7}) shows that the self-dual Yang-Mills equations
(\ref{3.7}) on $S^2\times H^2$ are equivalent to the BPS vortex-type equations  on $S^2$:
\begin{equation}\label{3.8}
F_{y\yb}=g_{y\yb}\,\left(\frac{1}{R_2^2}-\phi\bar\phi \right)\qquad\Leftrightarrow\qquad
\im\,F=\left(\frac{1}{R_2^2}-\phi\bar\phi \right)\,\omega_{S^2}^{}\ ,
\end{equation}
\begin{equation}\label{3.9}
\pa_{y}\phi + A_{y}\phi =0\quad\Leftrightarrow\quad\pa_A\phi =0\ ,
\end{equation}
where $\pa_A=\diff y(\pa_{y} + A_{y})$. Note that for the standard vortex equations instead of
eq.(3.9) one has $\pa_{\yb}\phi + A_{\yb}\phi =0$. This equation can be obtained if in (\ref{3.1})
one choose $\bar\b$ in the upper right corner and $-\b$ in the lower left corner (compact gauge
group SU(2)) but then in (\ref{3.8}) one will have $-1/R^2_2$ and such vortex-type equations will not
have solutions due to the Kazdan-Warner theorem~\cite{KW}.

Vortex number $N$ is defined as the first Chern number $c_1(E)$ of the bundle $E\to\C P^1$,
\begin{equation}\label{3.10}
N=c_1(E)=\frac{\im}{2\pi}\,\int_{S^2}^{}\, F\ .
\end{equation}
From (\ref{3.8}) it follows that
\begin{equation}\label{3.11}
\frac{\im}{2\pi}\,\int_{S^2}^{}F + \frac{1}{2\pi}\,\int_{S^2}^{}\phi\bar\phi\,\omega_{S^2}^{}=
\frac{1}{2\pi R_2^2}\,\int_{S^2}^{}\omega_{S^2}^{}= 2\left(\frac{R_1}{R_2}\right)^2\ ,
\end{equation}
and we obtain (cf.~\cite{Bradlow}) the inequality
\begin{equation}\label{3.12}
N \le 2\left(\frac{R_1}{R_2}\right)^2\ .
\end{equation}
For any $N\ge 0$ the condition (\ref{3.12}) can be satisfied for sufficiently large ratio $R_1/R_2$
and then  the moduli space of vortices on $S^2$ will be nonempty.

\bigskip

\noindent
{\bf Liouville-type equations on $S^2$.\ } Consider $N$-vortex solution $\phi=\exp(\sfrac12\,(u+\im\th ))$,
where $u$ and $\th$ are real-valued functions. Since $\phi$ can have zeros at $y_i\in\C P^1$, then $u(y)\to
-\infty$ as $y\to y_i$ and $\th(y)$ is a multi-valued function with ramification points at $y_i$. The
equation (\ref{3.9}) implies that
\begin{equation}\label{3.13}
A_y=-\pa_y\log\phi = -\sfrac12\pa_y(u+\im\th )\and A_{\yb}=\pa_{\yb}\log\bar\phi =\sfrac12\,\pa_{\yb}(u-\im\th)\ .
\end{equation}
Plugging (\ref{3.13}) into (\ref{3.8}), we obtain the Liouville-type equations on $S^2$,
\begin{equation}\label{3.14}
\pa_y\pa_{\yb}u=g_{y\yb}\left(\frac{1}{R^2_2} - e^u\right )\ ,
\end{equation}
away from the singularities of $u$.

Note that the sign on the right hand side of eq.(\ref{3.14}) with $g_{y\yb}$ given in (\ref{2.1}) is
opposite to the sign in the standard vortex equations on $S^2$. However, equations of type (\ref{3.14})
on a compact
Riemann surfaces (including $S^2$) were considered  by Kazdan and Warner~\cite{KW}. They have shown,
in particular, that equations
\begin{equation}\label{3.15}
\pa_y\pa_{\yb}u=\pm g_{y\yb}\left(\frac{1}{R^2_2} - e^u\right )
\end{equation}
have solutions for {\it both} signs in (\ref{3.15}) and equations
\begin{equation}\label{3.16}
\pa_y\pa_{\yb}u=\mp g_{y\yb}\left(\frac{1}{R^2_2} + e^u\right )
\end{equation}
have no solutions. These four cases exhaust possible  Liouville-type equations on $S^2$ with $R_2^2\ne\infty$.

Recall that equations (\ref{3.15}) can be obtained by the reduction of the self-dual Yang-Mills (SDYM) equations
from $S^2\times S^2$ to $S^2$ with gauge group SU(2) (lower sign) and from $S^2\times H^2$ to $S^2$ with
gauge group  SU(1,1) (upper sign). Similarly, equations (\ref{3.16}) correspond to the reduction of the
SDYM equations from $S^2\times H^2$ to $S^2$ with gauge group SU(2) (lower sign) and from $S^2\times S^2$
to $S^2$ with gauge group SU(1,1) (upper sign). Thus, only the gauge group SU(1,1) is allowed
for the considered case of the reduction $S^2\times H^2\to S^2$, and solutions of (\ref{3.14}) exist for any
$N\ge 0$.

If one considers the reduction of the SDYM equations from $S^2\times H^2$ to $H^2$, the allowed gauge group is
SU(2)~\cite{Popov}. In other words,  depending on a symmetry (SU(2)- or SU(1,1)-equivariance) imposed on
gauge fields, on $S^2\times H^2$ there exist solutions of the SDYM equations with gauge groups as SU(2) and
SU(1,1).

\section{Integrability of vortex equations on $S^2$}

\noindent
{\bf Integrable case.} We considered the BPS vortex-type equations (\ref{3.8})-(\ref{3.9}) and showed
their equivalence to the self-dual Yang-Mills equations (\ref{3.7}) on the manifold $M=S^2\times H^2$.
Note that for equal radii $R_1=R_2$ of $S^2$ and $H^2$ the scalar curvature (\ref{2.18}) of $M$ vanishes.
In this case the Weyl tensor for the manifold $M$ is self-dual~\cite{BLB}.

An important feature of K\"ahler manifolds $M$ with scalar curvature $R_M$ is that the so-called twistor
space $\Zcal$ of $M$ becomes a complex manifold  if $R_M=0$. Let us consider an open subset $\Ucal$ of
$M=S^2\times H^2$ with complex coordinates $y,z$. Then the twistor space of $\Ucal$ (i.e. the restriction
of $\Zcal$ to $\Ucal$) is diffeomorphic to $\Ucal\times\C P^1$, $\Zcal |_{\Ucal}\simeq \Ucal\times\C P^1$,
with a local complex coordinate $\lambda\in \C P^1\backslash\{\infty\}$ on the last factor. On $\Zcal$ there
is a distribution generated by three vector fields of type (0,1) closed under the Lie bracket. They have the
form (cf.~\cite{Popov})
\begin{equation}\label{4.1}
V_{\1}:=\tilde e_{\1}-\la\tilde e_{2}\ ,\quad V_{\2}:=\tilde e_{\2}+\la\tilde e_{1}
\and  V_{\3}=\pa_{\bar\la} \ ,
\end{equation}
where
\begin{equation}\label{4.2}
\tilde e_1=\rho_1^{-1}\left(\pa_{y} - (\pa_{y}\log\rho_1)\la\pa_{\la}\right) \ ,\quad
\tilde e_{\1}= \rho_1^{-1}\left(\pa_{{\yb}} + (\pa_{\yb}\log\rho_1)\la\pa_{\la}\right)\ ,
\end{equation}
\begin{equation}\label{4.3}
\tilde e_2=\rho_2^{-1}\left(\pa_{z} - (\pa_{z}\log\rho_2)\la\pa_{\la}\right) \ ,\quad
\tilde e_{\2}= \rho_2^{-1}\left(\pa_{\zb} +(\pa_{\zb}\log\rho_2)\la\pa_{\la}\right) \ .
\end{equation}
Recall that $\rho_1^2=g_{y\yb}$ and $\rho_2^2=g_{z\zb}$ are components of metrics on $S^2$ and
$H^2$; their explicit forms are given in section 2.

The vector fields (\ref{4.1}) define an almost complex structure $\J$ on $\Zcal$ such that
\begin{equation}\label{4.4}
\J(V_{\bar k}) = -\im V_{\bar k}
\end{equation}
for $k=1,2,3$.  For commutators of type (0,1) vector fields (\ref{4.1}) we have
\begin{equation}\label{4.5}
[V_{\1}, V_{\2}]=\la\rho_1^{-2}({\pa}_{y}\rho_1)V_{\1} + \la \rho_2^{-2}({\pa}_{z}\rho_2)V_{\2} +
2{{\la}^2}\left(\frac{1}{R_1^2}-\frac{1}{R_2^2}\right)V_{3}\ ,\quad
  {[V_{\1}, V_{\3}]}=0=[V_{\2}, V_{\3}]\ ,
\end{equation}
where $V_3=\pa_\la$ is the (1,0) vector field on $\Zcal$. Recall that for integrability of an almost
complex structure $\J$ on $\Zcal$ it is necessary and sufficient that the commutator of any two vector
fields of type (0,1) w.r.t. $\J$ is of type (0,1). For our case  we see from (\ref{4.5}) that $\J$ is
integrable - and $\Zcal$ is a complex manifold - if
and only if
\begin{equation}\label{4.6}
R_1=R_2\ ,
\end{equation}
i.e when the scalar curvature $R_M$ of the manifold $M=S^2\times H^2$ vanishes. In this case the bundle
$\Ecal\to M$ pulled back to the bundle $\hat\Ecal$ over the twistor space $\Zcal$ allows an integrable
holomorphic structure defined by a (0,1)-type connection along the vector fields (\ref{4.1}). The integrability
of this structure, $\Fcal^{0,2}=0$, is equivalent~\cite{Ward} to the self-duality equations on $M$.

\bigskip

\noindent
{\bf Lax pair.} For the case (\ref{4.6}) from (\ref{3.12}) we obtain the inequality
\begin{equation}\label{4.7}
N\le 2\ ,
\end{equation}
i.e. bundles $\hat\Ecal$ over $\Zcal$ with integrable holomorphic structures  describe configurations of $N=1$ and $N=2$
vortices on $S^2$. We emphasize that vortices exist for any $N>0$ but only for $N\le 2$ the vortex equations
(\ref{3.8})-(\ref{3.9}) appear from a Lax pair.

For presenting vortex equations on $S^2$ as an integrable system (for $R_1=R_2$) one should introduce
two linear equations (Lax pair) whose compatibility conditions will produce the vortex equations. For
that we introduce a (0,1) part $\hat\nabla^{0,1}$ of the covariant derivative $\hat\nabla$ on $\hat\Ecal$
by formulae
\begin{subequations}\label{4.8}
\begin{eqnarray}
\hat\nabla_{V_{\1}}^{}&\equiv&V_{\1}+\hat\Acal_{V_{\1}}:= \tilde e_{\1}-\la\tilde e_{2}+\Acal_{\1}-\la\Acal_2\ ,\\
\hat\nabla_{V_{\2}}^{}&\equiv&V_{\2}+\hat\Acal_{V_{\2}}:=\tilde e_{\2}+\la\tilde e_{1}+\Acal_{\2}+\la\Acal_1\ ,
\\
\hat\nabla_{V_{\3}}^{}&\equiv& V_{\3}+\hat{\Acal}_{V_{\3}}:=\pa_{\bar\la}\ ,
\end{eqnarray}
\end{subequations}
where $\hat\nabla_X$ denotes the covariant derivative along the vector field $X$.
The components
\begin{equation}\label{4.9}
\Acal_1=e^y_1\Acal_y\ ,\quad \Acal_{\1}=e^{\yb}_{\1}\Acal_{\yb}\ ,\quad \Acal_2=e^z_2\Acal_z\and
\Acal_{\2}=e^{\zb}_{\2}\Acal_{\zb}
\end{equation}
are easily extracted from (\ref{3.1}).

Let us now introduce a $2\times 2$ matrix $\psi = \psi (y, \bar y, z, \bar z, \la)$ which does not
depend on $\bar\la$ and consider two linear equations
\begin{subequations}\label{4.10}
\begin{eqnarray}
\hat\nabla_{V_{\1}}^{}\psi&:=&[\tilde e_{\1}+\Acal_{\1}-\la (\tilde e_{2}+\Acal_2)]\psi =0\ ,\\
\hat\nabla_{V_{\2}}^{}\psi&:=&[\la (\tilde e_{1}+\Acal_{1})+\tilde e_{\2}+\Acal_{\2})]\psi =0\ .
\end{eqnarray}
\end{subequations}
It is not difficult to check that the compatibility conditions of the linear equations (\ref{4.10}),
\begin{equation}\label{4.11}
\left([\hat\nabla_{V_{\1}}^{}, \hat\nabla_{V_{\2}}^{}] - \hat\nabla_{[V_{\1},V_{\2}]}^{}\right)\psi =0
\end{equation}
are equivalent to the vortex equations (\ref{3.8})-(\ref{3.9}) for $\Acal$ given in (\ref{3.1}).

Note that equations
\begin{equation}\label{4.12}
\hat\Fcal^{0,2}=0\quad\Leftrightarrow\quad \hat\Fcal (V_{\bar\imath}, V_{\bar\jmath})=[\hat\nabla_{V_{\bar\imath }}^{}, \hat\nabla_{V_{\bar\jmath}}^{}] - \hat\nabla_{[V_{\bar\imath},V_{\bar\jmath}]}^{}=0\ ,
\end{equation}
for $\hat\nabla_{V_{\bar\jmath}}^{}$ given in the first two formulae from (\ref{4.8}) can be imposed
even if an almost complex structure $\J$ on $\Zcal$ is not integrable, that is, the case when $R_1\ne R_2$.
Then equations (\ref{4.12}) define a pseudo-holomorphic structure~\cite{Br} on the bundle $\hat\Ecal\to\Zcal$.
These equations are again equivalent to the self-duality equations on $S^2\times H^2$ since
\begin{equation}\label{4.13}
\hat\Fcal(V_{\1}, V_{\2}) =
\Fcal_{\1\2}-\la\, (\Fcal_{1\1}+\Fcal_{2\2}) +\la^2\,\Fcal_{12}\ , \quad
\hat\Fcal(V_{\1}, V_{\3}) =0=\hat\Fcal(V_{\2}, V_{\3})\ ,
\end{equation}
where
\begin{subequations}\label{4.14}
\begin{eqnarray}
\Fcal_{\1\2}&=&e_{\1}\Acal_{\2}- e_{\2}\Acal_{\1} + [\Acal_{\1},\Acal_{\2}]=e_{\1}^{\yb}e_{\2}^{\zb}
\Fcal_{\yb\zb}\ ,\\
\Fcal_{12}&=&e_{1}\Acal_{2}- e_{2}\Acal_{1} + [\Acal_{1},\Acal_{2}]=e_{1}^{y}e_{2}^{z}
\Fcal_{yz}\ ,\\
\Fcal_{1\1}&=&e_{1}\Acal_{\1}- e_{\1}\Acal_{1} + [\Acal_{1},\Acal_{\1}]-
\r^{-1}_1(e_{\1}\r_1)\Acal_1 +  \r^{-1}_1(e_{1}\r_1)\Acal_{\1}=g^{y\yb}\Fcal_{y\yb}\ ,\\
\Fcal_{2\2}&=&e_{2}\Acal_{\2}- e_{\2}\Acal_{2} + [\Acal_{2},\Acal_{\2}]-
\r^{-1}_2(e_{\2}\r_2)\Acal_2 + \r^{-1}_2(e_{2}\r_2)\Acal_{\2}=g^{z\zb}\Fcal_{z\zb}\ .
\end{eqnarray}
\end{subequations}
After substituting SU(1,1)-equivariant gauge potential (\ref{3.1}), eqs.~(\ref{4.13}) reduce to the vortex
equations on $S^2$ having solutions with $N>2$. So, for $N>2$ vortex equations on $S^2$ do not appear as a
compatibility condition of a Lax pair but are derivable nevertheless from the self-dual Yang-Mills
equations similarly to vortex equations on Riemann surfaces with genus $g>1$,
where vortex equations were integrable only for $N\le 2(g-1)$~\cite{Popov}.

\section{Quiver vortex equations}

\noindent
Here and in section 6, we generalize the equations (\ref{3.8})-(\ref{3.9}) to the non-Abelian case
and describe relations between vortices on $S^2$ and instantons on the manifold $S^2\times\Sigma$,
where $\Sigma$ is a compact Riemann surface of genus $g>1$.

\noindent
{\bf Equivariant vector bundle.} Consider the manifold $M=\C P^1\times H^2$, where
$\C P^1\cong S^2$ is the Riemann sphere and $H^2$ is the unit disk described in section 2. Let
$\Ecal\to M$ be an SU(1,1)-equivariant rank-$k$ complex vector bundle, with the group SU(1,1) acting
trivially on $\C P^1$ and by isometry on $H^2=\ $SU(1,1)/U(1). Let $\Acal$ be a connection on $\Ecal$.
Imposing the condition of SU(1,1)-equivariance means that we should look for representations of the
group SU(1,1) on $\C^k$. Notice that for each positive integer $m$, the module
\begin{equation}\label{5.1}
\C^k = \mathop{\oplus}\limits^m_{i=0}\C^{k_i}\with \sum\limits^m_{i=0} k_i =k
\end{equation}
gives such a representation if $\C^{m+1}$ is an irreducible representation of SU(1,1). Let
\begin{equation}\label{5.2}
E= \mathop{\oplus}\limits^m_{i=0} E_i\quad\to\quad \C P^1
\end{equation}
be a rank-$k$ $\Z_{m+1}$-graded complex vector bundle over $\C P^1$ and $A^i$'s are connection forms
on the bundles $E_i\to\C P^1$. Then
\begin{equation}\label{5.3}
\Ecal = \mathop{\oplus}\limits^m_{i=0} \Ecal_i\with \Ecal_i=E_i\otimes L^{m-2i}\ ,
\end{equation}
where $L^{m-2i}=(L)^{\otimes (m-2i)}$ and the bundle $L\to H^2$ with a connection $a$ given in (\ref{2.13})
was introduced in section 2.

\bigskip

\noindent
{\bf Symmetric gauge potential and field strength tensor.} Similar to the compact SU(2) case~\cite{PS}, the
SU(1,1)-equivariant gauge potential $\Acal$ with values in End$\,\C^k$ decomposes into connections
$A^i\in u(k_i)$ on the complex rank-$k_i$ vector bundles $E_i\to\C P^1$ with $i=0,1,...,m$ and a multiplet
of scalar fields $\phi_{i+1}$ on $\C P^1$ with $i=0,1,...,m-1$ transforming in the bi-fundamental representation
$\C^{k_i}\otimes
(\C^\vee )^{k_i+1}$ of the group U($k_i$)$\times$U($k_{i+1}$), i.e. $\phi\in Hom(E_i, E_{i+1}$). Collecting
these Higgs fields into the upper triangular $k\times k$ complex matrix
\begin{equation}
\phi_{(m)}:=\begin{pmatrix}0&\phi_1&0&\dots&0\\0&0&\phi_2&\dots&0\\
\vdots&\vdots&\ddots&\ddots&\vdots\\0&0&0&\dots&\phi_m\\
0&0&0&\dots&0\end{pmatrix}\ ,
\label{5.4}
\end{equation}
we get
\begin{equation}
\label{5.5}
\Acal = A^{(m)}\otimes 1 + \Upsilon_{(m)}\otimes a +\frac{1}{\sqrt{2}}\,\phi_{(m)}\otimes\beta +
\frac{1}{\sqrt{2}}\,\phi_{(m)}^\+\otimes\bar\beta\ ,
\end{equation}
where
\begin{equation}
\label{5.6}
A^{(m)}:=\sum_{i=0}^mA^i\otimes\Pi_i\ , \quad \Upsilon_{(m)}:=\sum_{i=0}^m\,(m-2i)\ 1_{k_i}\otimes\Pi_i \ ,
\end{equation}
and $\Pi_i:E\to E_i$ are the canonical orthogonal projectors of rank 1, $\Pi_i\Pi_j =\delta_{ij}\Pi_j$, which
may be represented by diagonal $(m+1)\times (m+1)$ matrices $\Pi_i=(\delta_{ji}\delta_{li})_{j,l=0,1,...,m}$
of unit trace. Here $\b$ and $\bar\b$ are forms on $H^2$ of type (1,0) and (0,1) defined in
section 2.

The calculation of the curvature $\Fcal$ for $\Acal$ of the form (\ref{5.5}) yields
$$
\Fcal = \diff\Acal +\Acal\wedge\Acal = F^{(m)}\otimes 1 - \frac{1}{2}\Bigl(\frac{1}{R^2_2}\Upsilon_{(m)}
-[\phi_{(m)},\phi_{(m)}^\+]\Bigr )\b\wedge\bar\b
$$
\begin{equation}
\label{5.7}
+\frac{1}{\sqrt{2}}\Bigl(\diff\phi_{(m)}+[A^{(m)}, \phi_{(m)}]\Bigr )\wedge\b +\frac{1}{\sqrt{2}}\Bigl (
\diff\phi_{(m)}^\+ +[A^{(m)}, \phi_{(m)}^\+]\Bigr )\wedge\bar\b\ ,
\end{equation}
where $F^{(m)}=\diff A^{(m)} + [A^{(m)},A^{(m)}]$. The derivation of (\ref{5.7}) uses formulae (\ref{2.12}).

\bigskip

\noindent
{\bf Non-Abelian vortex equations on $\C P^1$.} Let us consider the self-dual Yang-Mills equations
$\ast\Fcal =\Fcal$ on $M$. In local coordinates on $M$ these equations have the form (\ref{3.7}). Substitution
of (\ref{5.7}) into the instanton equations on $M=S^2\times H^2$ reduce them to non-Abelian quiver vortex
equations (cf.~\cite{PS}-\cite{Popov})
\begin{equation}
\label{5.8}
\im\,F^{(m)}= \frac{1}{2}\Bigl(\frac{1}{R^2}\Upsilon_{(m)}-[\phi_{(m)}, \phi_{(m)}^\+]\Bigr )\,\omega_{S^2}\ ,
\end{equation}
\begin{equation}
\label{5.9}
\partial\phi_{(m)}+[A^{(m)}_{(1,0)}, \phi_{(m)}]=0\ ,
\end{equation}
where $\partial = \diff y\,\partial_y$ and $\omega_{S^2}$ is given in (\ref{2.2}). In terms of $(A^i, \phi_i)$
these equations have the form
\begin{equation}\label{5.10}
2\im\,F^{i}= \Bigl(\frac{m-2\im}{R^2}\,1_{k_i}+ \phi_{i}^\+\phi_{i}-
\phi_{i+1}\phi_{i+1}^\+\Bigr )\,\omega_{S^2}\ ,
\end{equation}
\begin{equation}
\label{5.11}
\partial\phi_{i+1}+A^{i}_{(1,0)}\phi_{i+1} - \phi_{i+1}A^{i+1}_{(1,0)}=0\ ,
\end{equation}
where $A^{i}_{(1,0)}=A^i_y\diff y, i=0,...,m$ and $\phi_0:=0=:\phi_{m+1}$. Finally, in local complex
coordinates on
$\C P^1$ we get
\begin{equation}\label{5.12}
2g^{y\bar y}F^{i}_{y\bar y}= \frac{m-2\im}{R^2}\,1_{k_i}+ \phi_{i}^\+\phi_{i}-
\phi_{i+1}\phi_{i+1}^\+\ ,
\end{equation}
\begin{equation}
\label{5.13}
\partial_y\phi_{i+1}+A^{i}_{y}\phi_{i+1} - \phi_{i+1}A^{i+1}_{y}=0\ .
\end{equation}

\bigskip

\section{Instantons with noncompact gauge groups}

\noindent
{\bf Riemann surfaces.} Recall that any simply connected Riemann surface $\Sigma$ of genus $g>1$ is
conformally equivalent to the unit disk $H^2=\ $ SU(1,1)/U(1). In other words, $H^2$ is a universal cover of
$\Sigma = H^2/\Gamma$, where $\Gamma$ is a Fuchsian group\footnote{It is a discrete subgroup of the group
SU(1,1)$\cong$SL(2,$\R$)}. Here, we will show that the ansatz (\ref{5.5}) can also be used on the manifold
$M=S^2\times\Sigma$, where $\Sigma$ is a compact Riemann surface of genus $g>1$.

The metric and the volume form on $\Sigma$ in local (conformal) coordinates $z, \bar z$ are given by
\begin{equation}\label{6.1}
\diff s^2_\Sigma = 2g_{z\zb}\,\diff z\diff\zb\and \omega_\Sigma =\im\,g_{z\zb}\,\diff z\wedge\diff\zb\ .
\end{equation}
Furthermore, for the nonvanishing components of the Christoffel symbols and the Ricci tensor we have
\begin{equation}\label{6.2}
\Gamma^z_{zz} = 2\,\pa_z\log\rho\quad\mbox{and}\quad
\Gamma^{\zb}_{\zb\zb} = 2\,\pa_{\zb}\log\rho\quad\mbox{with}\quad
\rho^2:=g_{z\zb}\ ,
\end{equation}
\begin{equation}\label{6.3}
R_{z\zb} = -2\,\pa_z\pa_{\zb}\log\rho =\varkappa\, g_{z\zb}\quad\Longrightarrow\quad
R^{}_{\Sigma} = 2 g^{z\zb}R_{z\zb}=2\varkappa\ ,
\end{equation}
where $R^{}_{\Sigma}$ is the constant scalar curvature of $\Sigma$.
The area of the Riemann surface with genus $g\ne 1$ is
\begin{equation}\label{6.4}
{\rm{Vol}}(\Sigma) = \int_{\Sigma}\omega_{\Sigma}^{} = \frac{4\pi}{\varkappa}\,(1-g)\ .
\end{equation}

Introducing forms $\b$ and $\bar\b$ of type (1,0) and (0,1) on $\Sigma$,
\begin{equation}\label{6.5}
\b :=\r\,\diff z\and \bar\b :=\rho\,\diff \zb\quad\Longrightarrow\quad \diff s^2_\Sigma =2\b\bar\b\ ,
\end{equation}
we obtain that
\begin{equation}\label{6.6}
 \diff\b = -2a\wedge\b\ ,\quad \diff\bar\b = 2a\wedge\bar\b\and
\diff a= \sfrac{1}{2}\,\varkappa\,\b\wedge\bar\b\ ,
\end{equation}
where
\begin{equation}\label{6.7}
2a=(\partial_z\log\r)\,\diff z - (\partial_{\zb}\log\r)\,\diff \zb
\end{equation}
is the Levi-Civita $u(1)$-connection on the tangent bundle $T\Sigma$ of $\Sigma$. Denoting the holomorphic part
$T^{1,0}\Sigma$ of $T\Sigma\otimes\C$ by $L^2$, we obtain the complex line bundle $L\to\Sigma$ with the
connection $a$. Finally, after choosing
\begin{equation}\label{6.8}
\varkappa = -\frac{1}{R^2}\ ,
\end{equation}
we see that $a,\b$ and $\bar\b$ in (\ref{6.6}) satisfy the same equations as forms in (\ref{2.12})
and therefore the ansatz (\ref{5.5}) on the manifold $\C P^1\times\Sigma$ yields to the curvature (\ref{5.7})
and to the quiver vortex equations (\ref{5.8})-(\ref{5.13}). That is why, in what follows we will consider our
gauge theory on the compact spaces $M=\C P^1\times\Sigma$.

\bigskip

\noindent
{\bf Reduction of the Yang-Mills functional.} The dimensional reduction of the Yang-Mills equations from
$\C P^1\times\Sigma$ to  $\C P^1$ can also be seen at the level of the Yang-Mills Lagrangian.
For simplicity, we consider the case $m=1$ for which the instanton equations on $\C P^1\times\Sigma$
are equivalent to the equations (\ref{3.8})-(\ref{3.9}) with $A=2A^0$, $\phi =\phi_1$ and $R=R_2$.
Substituting (\ref{5.5})-(\ref{5.7}) with $m=1$ into the standard Yang-Mills functional and performing
the integral over $\Sigma$, we arrive at the action
\begin{eqnarray}\nonumber
S&=&-\frac{1}{8\pi^2}\int_{S^2\times\Sigma}\tr\, (\Fcal\wedge *\Fcal )=
-\frac{1}{16\pi^2}\int_{S^2\times\Sigma}\diff^4 x\, \sqrt{\det (g_{\r\s})}\, \tr\,
(\Fcal_{\mu\nu}\Fcal^{\mu\nu})\\[4pt]
\label{6.9}
&=&(g-1)\frac{R^2}{4\pi}\int_{S^2}\omega_{S^2}^{}\,\Bigl\{(g^{y\yb})^2\,(F_{y\yb})^2 -
2g^{y\yb}(D_y\phi\, \overline{D_{y}\phi} + D_{\yb}\phi\,\overline{D_{\yb}\phi} )
+ \left(\frac{1}{R^2}-\phi\bar\phi \right)^2\Bigr\}\\[4pt]
\nonumber
&=&(g-1)\frac{R^2}{4\pi}\int_{S^2}\im\,\diff y\wedge\diff\bar y\,\Bigl\{g^{y\yb}\,\left(F_{y\yb}
+g^{y\yb}\,\left(\phi\bar\phi{-}\frac{1}{R^2}\right)\right)^2{-}4D_y\phi\, \overline{D_{y}\phi}\Bigr\}{+}
(g{-}1)\frac{\im}{2\pi}\int_{S^2} F\ ,
\end{eqnarray}
where $\m ,\n ,...=1,...,4,\ D_y=\partial_y+A_y$ and $D_{\yb}=\partial_{\yb}+A_{\yb}$. On solutions
($A, \phi $) of vortex equations (\ref{3.8})-(\ref{3.9}) this action coincides with $(g-1)N$, where
\begin{equation}\label{6.10}
N=\frac{\im}{2\pi}\,\int_{S^2}F=c_1(E)
\end{equation}
is the vortex number.

\bigskip

\noindent
{\bf Topological charges.} For self-dual gauge fields we have
\begin{equation}\label{6.11}
N_{inst}=-c_2(\Ecal)=-\frac{1}{8\pi^2}\,\int_{S^2\times\Sigma}\tr\, (\Fcal\wedge\Fcal )=
(g-1)\frac{\im}{2\pi}\,\int_{S^2}\, F= (g-1)\,N\ ,
\end{equation}
i.e. the instanton number $N_{inst}$ is proportional to the vortex number $N$. In the derivation of
(\ref{6.9})-(\ref{6.11}) it is assumed that $N\ge 0$.\footnote{For $N\le 0$ one should consider the
anti-self-dual Yang-Mills equations $\ast\Fcal = -\Fcal$ which reduce to anti-vortex equations.} From
(\ref{6.9}) we see that due to noncompactness of the gauge group SU(1,1) the energy density
for vortices is not positive definite but for $(A,\phi )$ satisfying the BPS vortex equations
(\ref{3.8})-(\ref{3.9}) the action $S$ coincides with the topological invariant $(g-1)c_1(E)=-c_2(\Ecal)$.
Thus, by solving equations (\ref{5.8})-(\ref{5.9}) on $\C P^1$ one can obtain instantons on
$\C P^1\times\Sigma$ with noncompact gauge group and the topological charge
$N_{inst}=(g-1)\,N$.

\bigskip

\noindent
{\bf Acknowledgments.} I would like to thank Olaf Lechtenfeld for useful discussions and the Institute
for Theoretical Physics of Leibniz Universit\"at Hannover, where this work was completed, for hospitality.
This work was partially supported by the Alexander von Humboldt Foundation.

\end{document}